\documentclass[aps,prl,preprint,groupedaddress]{revtex4-1}

\usepackage{graphicx,amssymb,amstext,amsmath} 
\usepackage{color}

\begin{document}

\title{Influence of Fluctuating Membranes on Self-Assembly of Patchy Colloids}

\author{Richard Matthews}
\author{Christos N. Likos}%

\affiliation{%
Faculty of Physics, University of Vienna, Boltzmanngasse 5, A-1090 Vienna, Austria
}%

\date{\today}

\begin{abstract}
A coarse-grained computational model is used to investigate the effect of a fluid membrane on patchy-particle assembly into biologically-relevant structures motivated by viral cores and clathrin. For cores, we demonstrate a non-monotonic dependence of the promotion of assembly on membrane stiffness. If the membrane is significantly deformable, cores are enveloped in buds, although this effect is suppressed for very flexible membranes. In the less deformable regime, we observe no marked enhancement for cores, even for strong adhesion to the surface. For clarthrin-like particles, we again observe the formation of buds, whose morphology depends on membrane-flexibility.
\end{abstract}


\maketitle


In self-assembly, the interactions between a collection of components guide them to spontaneously form an ordered structure~\cite{whitesides}. Biological self-organization happens within cells, from which all living organisms are composed. Cells are all bounded by a membrane, as are many sub-cellular structures. Thus many self-assembly processes are membrane-influenced. Membranes themselves are also self-assembled, primarily as a lipid bilayer~\cite{israelachvili}. We focus, however, on structures assembled only from proteinaeous sub-units, particularly viruses and clathrin. 

The genome of a virus is contained in a core or capsid, a typically mono-disperse shell, assembled from individual protein complexes. Often the shells are approximately spherical, with many having icosahedral symmetry~\cite{baker}. Viruses are divided into enveloped and non-enveloped types, depending on whether the core is surrounded by a membrane. The envelope in the former group is acquired through budding~\cite{cann}. For both enveloped~\cite{gelderblom,ono,miyanari,shavinskaya,forsell,ng} and non-enveloped~\cite{simon,siegel,bravo} viruses there is abundant evidence of membrane influence on core assembly. Clathrin, on the other hand, is intrinsically linked to membranes: its main function is the formation of coated vesicles for intra-cellular protein transport~\cite{brodsky}. Its three-legged shape allows a collection of individual units to form structures that range from extended hexagonal sheets to closed cages, which always include 12 pentagonal, in addition to different numbers of hexagonal, faces~\cite{fotin}. Assembly is nucleated on cellular membranes by adaptors, protein complexes which bind the lattice to the membrane. Hexagonal sheets on membranes are observed~\cite{heuser} and coated vesicles form through budding~\cite{brodsky}.

Experimentally, the reversible disassembly and reassembly of viral capsids in solution may be triggered by raising and lowering the pH~\cite{fraenkel}, allowing {\it in vitro} experiments of bulk assembly, which is observed, for example, by light scattering~\cite{mukherjee} or electron microscopy~\cite{sorger}. Similar experiments with clathrin~\cite{zaremba} observed bulk assembly into cage structures, finding them to be much more homogeneous when adaptor proteins are present.

Much theoretical work on biological bulk self-assembly has used patchy-particle models. Patchy-particles have discrete, attractive interaction sites on their surface and are very versatile in terms of the range of structures that may be assembled~\cite{bianchi}. The main focus has been on the assembly of mono-disperse viral capsids~\cite{rapaport, nguyen, hagan, wilber, johnston}, with simulations reproducing key characteristics such as a lag time, hysteresis and partial capsid formation at high concentrations. Simulations also give more detailed, experimentally inaccessible information about assembly dynamics. A similar coarse-grained simulation approach was also applied to clathrin assembly~\cite{otter,otter2}.

Previous applications of coarse-grained models to the effect of membranes on self-assembly are limited, although the effect of rigid templates has been considered~\cite{williamson}. Aggregation of isotropic spherical particles, on fluid~\cite{saric}, and polymerized~\cite{saric2011}, membranes was studied. More detailed models were also applied to self-assembly within a lipid bilayer~\cite{bond,khalfa}. Although not considering assembly, simulations of particles attracted to a membrane also saw budding~\cite{reynwar}. An alternative continuum approach found that self-assembly induced budding is controlled by interaction strengths and rigidity~\cite{zhang:pre:2008}.

Given the evidence of the influence of membranes on the self-assembly of biological structures, it is important to explore the generic physics that plays a role in such systems. Although biological detail is undoubtedly important, we choose rather to investigate coarse-grained models that share salient features with viral cores and clathrin. We focus on a range of interaction strengths that cover the crossover to assembly in the bulk, as well as that from unbound to membrane-bound structures. We correspondingly choose bending rigidities within a range whose lowest end gives membranes that are easily deformed, and whose highest end gives ones that may not be deformed, by the assembled structures. Here we focus on equilibrium, postponing dynamics to a later work.


Our model comprises $N_{SU}$ assembling sub-units plus the membrane. The former are modeled as spherical patchy-particles with a Kern-Frenkel potential~\cite{bianchi}, similarly to previous work~\cite{wilber} but modified suitably such that its first derivative is continuous. The membrane is represented using a dynamically triangulated surface model~\cite{gompper}: $N_{mem}$ particles connected with $3N_{mem}$ bonds form a network of $N_{tri} = 2N_{mem}$ triangles. We sample using Monte Carlo (MC) simulations~\cite{frenkel}, performed in a periodic rectangular box of sides $L_x$, $L_y$ and $L_z$. The membrane's projection completely covers the box in the $xy$-plane, connecting to itself across the boundaries. To apply no external tension~\cite{schmid} we allow $L_x = L_y$ to vary, whilst also adjusting $L_z$ to keep to volume, $V = L_x L_y L_z$, fixed. The MC moves used do not allow the membrane topology to change. Sub-units interact with the membrane from both sides but are only attracted to one side. Quantities are given in units of the thermal energy, $k_B T$, or the typical length of a membrane bond, $l$.

Interactions between sub-units ($ss$) and between sub-units and membrane particles ($ms$) are of a Lennard-Jones type (see appendix). An orientational dependence of the attractive part creates patches. The parameter $\theta_0$ defines the maximum angular deviation of the patch position from the particle-to-particle vector before the attractive interaction decreases. It is chosen such that, for a given pair of sub-units, only one pair of $ss$ patches can interact at once. For $ss$ interactions, twisting of sub-units around an interacting patch is also penalized, mimicking the torsional constraints in protein-protein interactions~\cite{wilber}. The $ss$ patches are evenly spaced around the particle, with the $ms$ patch lying on the axis of rotational symmetry. The sub-units have a size of $\approx 2.5$ and we choose $\theta_0 = \pi/4$  for the $ms$ interactions. This gives a relatively wide $ms$ patch, so that a sub-unit typically interacts with many membrane particles so seeing a smooth surface. The minima of the $ss$ and $ms$ interactions are $-\epsilon_{ss}$ and $-\epsilon_{ms}$. A pair of sub-units are defined to be bonded if their interaction energy is $<-0.25\epsilon_{ss}$. The bending stiffness of the membrane is set by $\lambda_b \simeq \kappa$, the bending rigidity of the membrane (see~\cite{gompper} and appendix).


We choose two different parameters sets. For our core model, $N_{SU} = 12$. Sub-units have five $ss$ patches with $\theta_0 = 0.2$, giving a similar patch width to the optimum in Ref.~\cite{wilber}. If the twelve sub-units are placed on the vertices of an icosahedron, they may be aligned with the $ms$ patches pointing outwards and every $ss$ patch pointing directly at a patch on a neighboring sub-unit, bonding with it. Here, $V =  1.07\times10^4$ and $N_{mem} = 576$. In our clathrin-like model $N_{SU} = 36$ and sub-units have three $ss$ patches each with $\theta_0 = 0.3$. The patches are wider to allow for a range of curvatures. Following Ref.~\cite{otter2}, the $ss$ patches make an angle of $(79/180) \pi$ to the $ms$ patch so that, if a closed cage is formed, the $ms$ patches point inwards. Here, $V =  2.15\times10^4$ and $N_{mem} = 1156$. Membrane sizes were chosen, using preliminary runs, to give plenty of area to cover assembled structures, with the $V$ chosen to allow large membrane deformations. Qualitative results were not sensitive to $V$.

The main connection to biological systems is that the interactions drive our models to form similar structures. The core sub-units resemble intermediate capsomers in the assembly of a $T = 1$ capsid, the smallest virus structure with icosahedral symmetry~\cite{wilber}. In reality, enveloped viruses are larger. In our clathrin-like model sub-units are considered equivalent to one clathrin, with each patch representing a leg. This is a simplification in that, in structure formation, multiple legs of different clathrin lie along each other. We neglect adaptor proteins~\cite{brodsky}.


Efficient sampling of our system must overcome a number of issues: free-energy barriers between assembled and disassembled states; the importance of collective motion for membrane relaxation; large times to find target structures. MC allows us to combine different approaches that address these problems, specifically Aggregate Volume Bias (AVB) moves~\cite{chen_AVBMC}, Hybrid Monte Carlo (HMC) moves~\cite{mehlig} and Multicanonical Parallel-Tempering (MPT)~\cite{faller}. The AVB moves, which shift sub-units directly between non-bonded and bonded states, as well as displacing bonded clusters onto or off the membrane, allow target structures to be found very quickly. HMC, which uses molecular dynamics (MD) trajectories to create trial states, captures collective motion. Finally, the free-energy barrier problem is ameliorated through the use of MPT, involving parallel tempering swaps in two dimensions, $\epsilon_{ss}$ and $
\epsilon_{ms}$. The further addition of a one-dimensional biasing potential, $w(E_{ss} /\epsilon_{ss})$, constructed iteratively during initialization, increases the swap acceptance rate. $E_{ss}$ is the total interaction energy between all sub-units. We found that the HMC acceptance rate is significantly improved by constructing $w(E_{ss}/\epsilon_{ss})$  as a continuous differentiable function and including the resulting forces in the MD integration~\cite{bartels}.

\begin{figure}[ht]
\begin{center}

\includegraphics[scale=0.6]{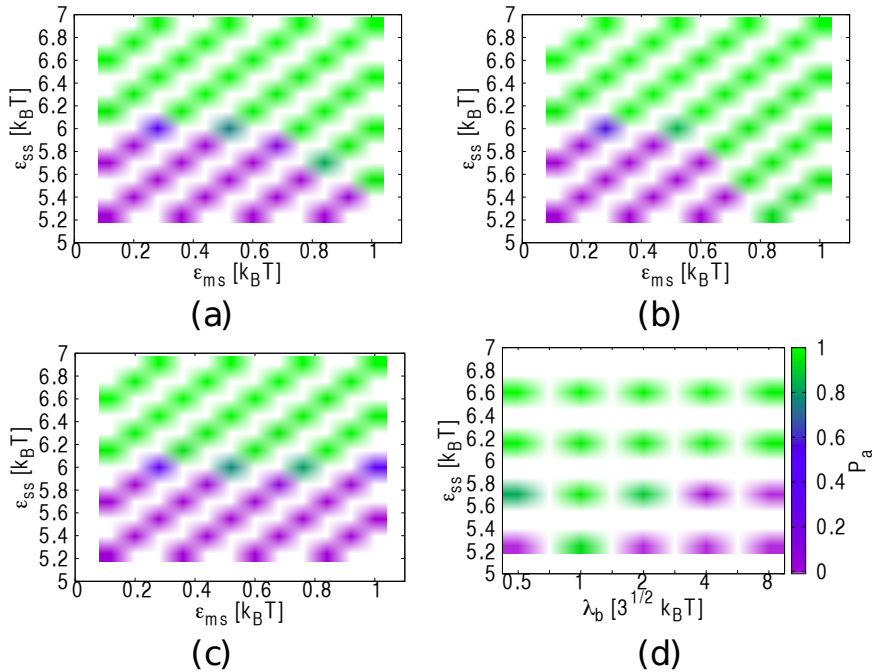}

\caption{\label{fig:P_assemb} Probability of finding a correctly assembled icosahedral core in a simulation with 12 sub-units as a function of $\epsilon_{ss}$ and $\epsilon_{ms}$ for different $\lambda_b$: (a) $\sqrt{3}/2 $ (b) $\sqrt{3} $ (c) $8 \sqrt{3}$. Results for $2\sqrt{3}$ and $4\sqrt{3}$ are shown in the appendix. (d) As a function of $\epsilon_{ss}$ and $\lambda_b$ for $\epsilon_{ms} = 0.84$.}
\end{center}
\end{figure}

We define a core to be assembled if all twelve sub-units are in a cluster and each makes five bonds. In Fig.~\ref{fig:P_assemb} we plot the probability of finding an assembled core, $P_{a}$, as a function of $\epsilon_{ss}$ and $\epsilon_{ms}$ for a range of $\lambda_b$ between $\sqrt{3}/2 $ and $8\sqrt{3} $. Our chosen $\epsilon_{ss}$ range covers the crossover from $P_{a} \approx 0$ to $P_{a} \approx 1$ for a bulk system with the same free assembly volume. For all $\lambda_b$, we observe that, for the lowest $\epsilon_{ms},$ this crossover occurs at about the same $\epsilon_{ss}$ as in the no-membrane system (see appendix). 

\begin{figure}[ht]
\begin{center}

\includegraphics[scale=0.6]{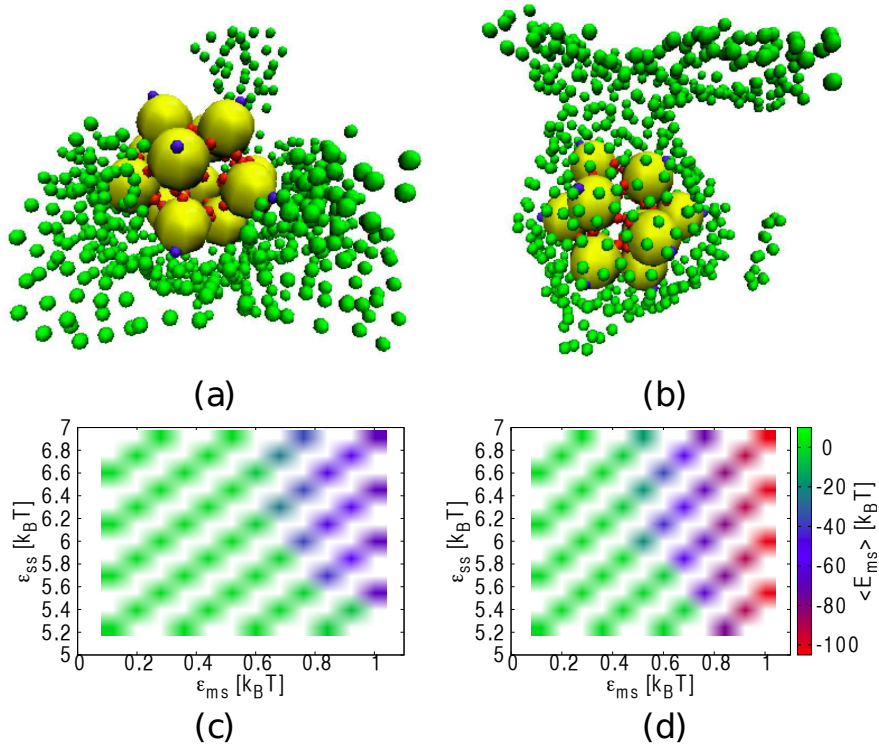}

\caption{\label{fig:core_snapshots} Typical configurations for an assembled core strongly attracted to the membrane, taken from simulations with $\epsilon_{ss} = 5.55 $ and $\epsilon_{ms} = 1$ for $\lambda_b$: (a) $\sqrt{3}/2 $ (b) $\sqrt{3} $. Membrane particles are shown in green and sub-units in yellow. The positions (not extents) of the $ss$ patches are shown in red and the $ms$ patches in blue. The average total membrane-sub-unit interaction energy, $\langle E_{ms} \rangle$, as a function of $\epsilon_{ss}$ and $\epsilon_{ms}$ is plotted for the same $\lambda_b$: (c)  $\sqrt{3}/2 $ (d)  $\sqrt{3} $.}
\end{center}
\end{figure}

For more deformable membranes, as $\epsilon_{ms}$ is increased, assembly occurs at lower $\epsilon_{ss}$.  This enhancement depends non-monotonically on $\lambda_b$, see Fig.~\ref{fig:P_assemb}(d), occurring over a larger area of parameter space for $\lambda_b = \sqrt{3}$ than for $\lambda_b = \sqrt{3}/2$, but then reducing and disappearing as $\lambda_b$ is increased further. For lower $\lambda_b$, and high $\epsilon_{ms}$, the membrane tends to envelop the sub-units. In Fig.~\ref{fig:core_snapshots}, typical configurations observed for $\lambda_b = \sqrt{3}/2 $ and $\lambda_b = \sqrt{3} $ with an assembled core attached to the membrane are shown. Interestingly, whilst for $\lambda_b =  \sqrt{3} $ this envelopment is almost complete, forming a bud, for $\lambda_b = \sqrt{3}/2 $ it is only partial. In Figs.~\ref{fig:core_snapshots}(c) and (d) we plot the average of the total membrane-sub-unit interaction energy, $\langle E_{ms} \rangle$, as a function of $\epsilon_{ss}$ and $\epsilon_{ms}$ for the same $\lambda_b$, confirming that for $\lambda_b = \sqrt{3}/2 $ the membrane envelops the sub-units less: for $\lambda_b =  \sqrt{3} $ the minimum of $\langle E_{ms} \rangle$ is $\approx -100 $, whilst for $\lambda_b =  \sqrt{3}/2 $ it is $\approx -70 $. The lowest $\langle E_{ms} \rangle$ are strongly correlated with envelopment in buds.

For $\lambda_b = 2\sqrt{3}$, some configurations with similar envelopment as for $\lambda_b = \sqrt{3}$ are seen but, for higher $\lambda_b$ only some deformation, not full envelopment, is seen (see appendix). The lack of an enhancement of assembly in this regime, despite strong attractions to the membrane, is in contrast to the case of extended crystals, where structures grow near attractive walls even if the bulk is fluid~\cite{teeffelen}.

In the bulk, the probability of assembly is determined by whether the attractions are sufficient to overcome the associated entropy loss. The attraction of sub-units to the membrane confines them, reducing this entropy loss, which may promote assembly. If the free energy gain in forming the core is sufficient to overcome the bending energy, as well as the entropic cost of binding to the core, the assembled structure may form a bud. Budding is not necessary for assembly promotion but the membrane stiffness with the most budding also has the most promotion. The non-monotonic rigidity dependence may be due to budding suppression: for low stiffness by membrane entropy and for high stiffness by bending energy. For our cores, changing the membrane stiffness and attraction changes the probability of forming one specific structure. For isotropic particles, in contrast, altering these parameters may lead to qualitatively different structures~\cite{saric,saric2011}.

Whilst our range of bending rigidities approximately overlaps with that expected for biological membranes ($\approx 2.5 - 25$ \cite{evans:arbbs:94}), those where we see budding are somewhat 
on the lower side of this range ($\lambda_b \leq 2\sqrt{3}$). This discrepancy may well arise from the coarse-grained nature of our model and particularly from the relatively small number of sub-units in our cores: assuming the free energy gain is proportional to the number of sub-units forming them, smaller structures will be less able to deform the membrane into a given shape.

\begin{figure}
\begin{center}

\includegraphics[scale=0.6]{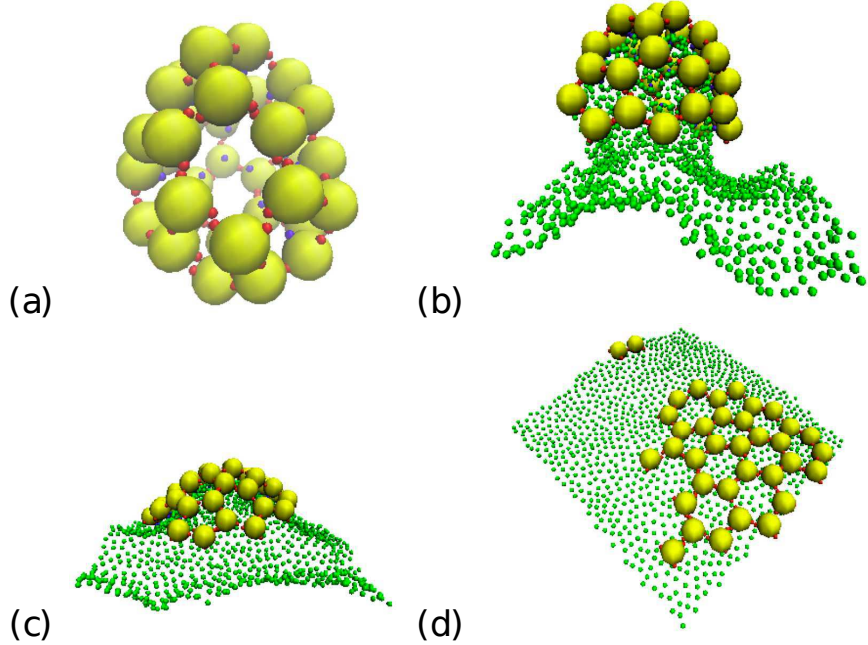}

\caption{\label{fig:clath_snapshots} Typical configuration from simulations with the clathrin-like model: (a) $\epsilon_{ss} = 12 $, without membrane. (b)  $\epsilon_{ss} = 12 $,  $\epsilon_{ms} = 1$,  $\lambda_b = \sqrt{3} $  (c)  $\epsilon_{ss} = 12 $,  $\epsilon_{ms} = 1$,  $\lambda_b = 2\sqrt{3} $. (d)  $\epsilon_{ss} = 12 $,  $\epsilon_{ms} = 1$,  $\lambda_b = 8\sqrt{3} $. Coloring as in Fig.~\ref{fig:core_snapshots}.}
\end{center}
\end{figure}

For the clathrin-like model, the structures formed are typically poly-disperse, see Fig.~\ref{fig:clath_snapshots}, and we use a standard measure of asphericity, $\Delta$, to investigate their shape.  $\Delta$ (see~\cite{aronovitz} and appendix) takes values between 0 and 1, with 1 corresponding to a shape with spherical symmetry and 0 corresponding to a non-spherical, oblate or prolate shape.

\begin{figure}
\begin{center}

\includegraphics[scale=0.6]{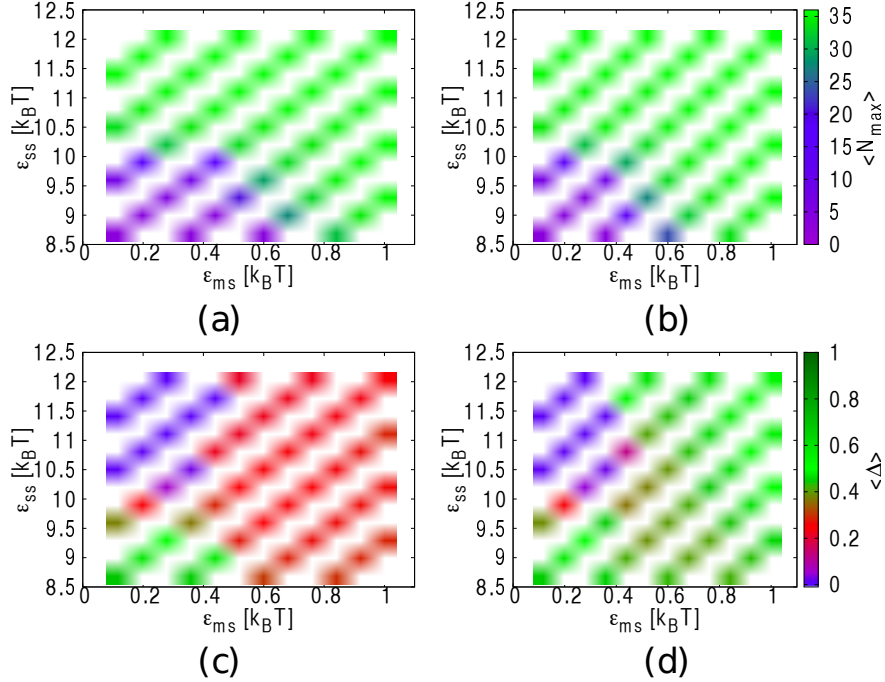}

\caption{\label{fig:clath_N_max} Clathrin-like model: average number of sub-units in the largest cluster, $\langle N_{max} \rangle$, as a function of $\epsilon_{ss}$ and $\epsilon_{ms}$ for $\lambda_b$: (a) $\sqrt{3}/2 $ (b) $8 \sqrt{3} $. Average asphericity of the largest cluster, $\langle \Delta \rangle$, as a function of  $\epsilon_{ss}$ and $\epsilon_{ms}$ for $\lambda_b$: (c) $2\sqrt{3} $ and (d) $8\sqrt{3} $.}
\end{center}
\end{figure}

We first focus on the average of the number of sub-units in the largest bonded cluster, $\langle N_{max} \rangle$. We consider the same range of $\epsilon_{ms}$ as for the core model and choose the $\epsilon_{ss}$ range so that for the no-membrane system it covers the crossover from small clusters of a few sub-units to most of the 36 sub-units being in one cluster (see appendix). For higher $\epsilon_{ss}$, without a membrane, the sub-units are observed to form closed cages, see Fig.~\ref{fig:clath_snapshots} (a). We observe that the sub-unit bonds form 5 and 6 member closed rings on the cage surfaces but the shape of the ``faces'' they enclose deviate significantly from pentagons or hexagons, being not generally flat, and the cages, whilst qualitatively similar, are not generally of the form of the structures observed for clathrin~\cite{fotin}. The key difference may be that when two clathrin bond their legs lie along each other, which will result in a greater flexibility to tilt the symmetry axes of the two clathrin than to rotate around the symmetry axes. In our clathrin-like model, however, a bond has equal flexibility for both such deformations.

In Fig.~\ref{fig:clath_N_max} (a) and (b) we plot $\langle  N_{max} \rangle$ as a function of $\epsilon_{ss}$ and $\epsilon_{ms}$ for $\lambda_b = \sqrt{3}/2$ and $8\sqrt{3}$. Looking at Fig.~\ref{fig:clath_N_max}(a) we see that, for the most flexible membrane, there is a similar enhancement of assembly for higher $\epsilon_{ms}$ as for cores. However, as shown in Fig.~\ref{fig:clath_N_max}(b), in contrast to cores, the enhancement remains as $\lambda_b$ is increased. The results for all intermediate $\lambda_b$ were very similar (see appendix). This is due to the ability of the clathrin-like sub-units to form structures with a range of curvatures: as depicted in Fig.~\ref{fig:clath_snapshots}(b), at lower $\lambda_b$ the sub-units form roughly spherical structures which enclose membrane buds. Indeed, at $\lambda_b = \sqrt{3}/2$, these are often nearly closed, with the membrane in the bud connected to the rest by a very narrow neck (see appendix). At intermediate $\lambda_b$, we observe the formation of more open, curved structures, or pits, on the membrane, see Fig.~\ref{fig:clath_snapshots}(c), whereas at high $\lambda_b$ the sub-units form extended, approximately flat structures lying on the membrane surface, as shown in Fig.~\ref{fig:clath_snapshots}(d).

In Fig.~\ref{fig:clath_N_max}(c) and (d) we plot the average of the asphericity of the largest sub-unit cluster, $\langle \Delta \rangle$, for $\lambda_b = 2\sqrt{3}$ and $8\sqrt{3} $. For low $\epsilon_{ms}$ and  higher $\epsilon_{ss}$ there is a region for all $\lambda_b$ where $\langle \Delta \rangle \approx 0$, corresponding to closed cage structures, not attached to the membrane. For higher $\epsilon_{ms}$, non-closed structures bound to the membrane are formed. At lower $\lambda_b$ these remain somewhat spherical, whilst for $\lambda_b = 8\sqrt{3} $ they are a lot less so.

To summarize, we described a simple coarse-grained model to simulate the effect of a membrane on the assembly of proteinaceous sub-units. We used this model to investigate the assembly of structures that share key features with viral cores and clathrin. In both cases we found that attraction to the membrane may enhance assembly in regions without bulk assembly. For cores, this effect shows an interesting non-monotonic dependence on membrane rigidity, being reduced for very deformable membranes and disappearing for the stiffest, in contrast to extended crystalline structures with attractive walls. For clathrin-like particles, the promotion of assembly persists for less deformable membranes. The difference to cores is due to the ability of clathrin-like particles to form structures with different curvatures. Furthermore, we observed the formation of biologically relevant buds for both cores and clathrin-like particles. In the case of cores, we found that these do not occur if the membrane is very flexible, whilst for clathrin-like particles their morphology depends on membrane-flexibility.

The formation of buds on membranes is crucial in various biological processes, for example endocytosis, in which, in some organisms, clathrin plays an important role. Endocytosis is a complex process involving the collaborative binding of variety of proteins to the membrane~\cite{liu}. The demonstration of bud-formation through assembly in our simulations opens the possibility that simple, patchy-particle models could capture basic features of such processes, giving new insight. The effects described might also be experimentally observed by mixing patchy colloids \cite{ilona:mrc:2010,granick:nature:2011} with giant vesicles~\cite{discher:science:99,keller:bj:2003}, whose bending rigidity~\cite{helfrich:pra:91} lies well within the range considered. More generally, our results clearly demonstrate that membranes can have a profound impact on self-assembly and will hopefully stimulate further study in this direction. In future it will be interesting to investigate the dynamics of membrane-influenced assembly.

This work was supported by the Austrian Science Fund (FWF): M1367. Snapshots were created using VMD~\cite{humphrey}. The computational results presented have been achieved in part using the Vienna Scientific Cluster (VSC).


%

\clearpage
\section{APPENDIX}

We give additional details of our simulation model. We also include a number of extra plots and snapshots. Parameters and variables are as defined in the main text.

\subsection{Model Details}

For the interactions between membrane particles we use smooth potentials that are also appropriate for molecular dynamics~\cite{noguchi}. Bonded membrane particles interact via
\begin{equation}
U_{bond}(r_{ij}) = 
\left\{
\begin{array}{l}
0\\
\hspace{1.2 in} \mathrm{for} \; r_{ij} \le 1.15l, \\
(80 k_BT)\exp[1/(1.15l - r_{ij})] / (1.33l - r_{ij})\\
\\
\hspace{1.2 in} \mathrm{for} \;  1.15l < r_{ij}  < 1.33l, \\
\infty \\
\hspace{1.2 in} \mathrm{for} \; r_{ij} \ge 1.33l, \\
\end{array}
\right.
\end{equation}
with $r_{ij} = |\mathbf{r}_{ij} | =  | \mathbf{r}_j - \mathbf{r}_i | $, where $\mathbf{r}_{i}$  is position of particle $i$. All pairs of membrane particles experience an excluded volume potential
\begin{equation}
U_{EV}(r_{ij}) = 
\left\{
\begin{array}{l}
\infty \\
\hspace{1.2 in} \mathrm{for} \; r_{ij} \le 0.67l, \\
(80 k_BT)\exp[1/(r_{ij} - 0.85l)] / (r_{ij} - 0.67l)\\
\\
\hspace{1.2 in} \mathrm{for} \; 0.67l < r_{ij} < 0.85l,\\
0\\
\hspace{1.2 in} \mathrm{for} \; r_{ij} \ge 0.85l.\\
\end{array}
\right.
\end{equation}
The minimum distance between any two membrane particles is $0.67l$ and the maximum bond length is $1.33l$. 

A unit normal vector is associated with each membrane triangle. Each bond forms the side of two different neighboring triangles. Membrane fluidity is included using MC moves that attempt to remove a given bond and create a new one between the two vertices of its neighboring triangles that were not connected by the original. During this procedure the direction of the normals is always maintained such that if the membrane were in a flat configuration all normals would point in the $+z$-direction. The bending stiffness of the membrane is controlled by including a potential $U_{bend} = \lambda_b (1 - \mathbf{n}_i \cdot \mathbf{n}_j)$ for each bond, where $ \mathbf{n}_i$ and $ \mathbf{n}_j$ are the unit normal vectors of the two triangles neighboring the bond and $\lambda_b$ is an energy. This form for the bending energy has the weakness that the effective bending rigidity is shape dependent. We have also performed simulations using an alternative that does not share this deficiency~\cite{gompper}. We observed qualitatively similar results but found simulation times were typically increased by about a factor of two. The total membrane area, $A$, is constrained with a harmonic potential, $U_{area} = (k_BT)(A - A_0)^2$, where $A_0 = (\sqrt{3}/4) l^2 N_{tri}$, in the Hamiltonian.

For the case where particles $i$ and $j$ ($i \ne j$) are both sub-units ($ss$) or where one is a sub-unit and one a membrane particle ($ms$) the interaction is of the following form, 
\begin{eqnarray}
\nonumber U_{ij}&=& \gamma_{area} \left[U_{WCA}(r_{ij}) + \gamma_{side} \gamma_{orient} U_{att}(r_{ij}) \right]
\\\nonumber U_{WCA}(r)&=& 
\left\{
\begin{array}{l}
4\epsilon\left[\left(\frac{\sigma}{r}\right)^{12}-\left(\frac{\sigma}{r}\right)^{6} + \frac{1}{4} \right]\\
\hspace{1.2 in} \mathrm{for} \; r < r_t,  \\
0\\
\hspace{1.2 in} \mathrm{for} \; r \ge r_t, \\
\end{array}
\right.
\\U_{att}(r)&=& 
\left\{
\begin{array}{l}
-\epsilon\\
\hspace{1.2 in} \mathrm{for} \; r < r_t,\\
4\epsilon\left[\left(\frac{\sigma}{r}\right)^{12}-\left(\frac{\sigma}{r}\right)^{6} \right]\\
\\
\hspace{1.2 in} \mathrm{for} \; r_t  \le r \le r_s,\\
a(r - r_c)^2 + b(r - r_c)^3\\ 
\\
\hspace{1.2 in} \mathrm{for} \; r_s \le r \le r_c,\\
0 \\
\hspace{1.2 in} \mathrm{for} \; r \ge r_c,\\
\end{array}
\right.
\label{eq:EV_pot}
\end{eqnarray}
where $r_t = 2^{1/6}\sigma$, $r_s = (\frac{26}{7})^{1/6}\sigma$, $r_c = \frac{67}{48}r_s$, $a = - \frac{24192}{3211}\frac{\epsilon} {r_s^2}$ and $b = -\frac{387072}{61009}\frac{\epsilon}{r_s^3}$. The form of $U_{att}(r)$ in the range $ r_s \le r \le r_c$ is a polynomial interpolation used to avoid a jump in the potential or its first derivative at the cut-off~\cite{bordat}. The dimensionless factors $\gamma_{area}$, $\gamma_{side}$ and $\gamma_{orient}$ take different forms for $ss$ and $ms$ interactions. For all simulations reported, we set the length scale, $\sigma$, for $ss$ interactions to $\sigma_{ss} =  2.5 l$ and similarly set $\sigma_{ms} = (1/2)(\sigma_{ss} + l)$. The corresponding energy scales, $\epsilon$, denoted $\epsilon_{ss}$ and $\epsilon_{ms}$ are varied.

Since the membrane bonds have a relatively broad, flat minimum the membrane particles would tend to be locally compressed when an attractive sub-unit is close. For $ms$ interactions the $\gamma_{area}$ factor is used to counter-act this by making the interaction proportional to the area that the membrane particle represents: $\gamma_{area} = A_{neigh}/(N_{neigh}A_{tri}$), where $N_{neigh}$ is the total number of triangles that have the membrane particle as a vertex, $A_{neigh}$ is their total area and $A_{tri} = A_0 / N_{tri}$. For $ms$ interactions $\gamma_{side}$ is used to make only one side of the membrane attract sub-units: it takes a value of $1$ if the sub-unit is ``above'' the membrane and $0$ if it is ``below''. A sub-unit is determined to be ``above'' or ``below'' by finding the closest point  on the membrane in the $z$-direction. If the normal of the triangle enclosing the closest point makes an angle of less than $\pi / 2 $ with the vector from the closest point to sub-unit then the sub-unit is ``above'' the membrane, otherwise it is ``below''. For $ss$ interactions $\gamma_{area} = \gamma_{side} = 1$. 

The centres of the attractive patches are defined by unit vectors. The width of the patches is determined by $\gamma_{orient}$, which is a product of functions of the form~\cite{miller}:
\begin{equation}
F(\phi; \phi_0, \phi_1) = 
\left\{
\begin{array}{l}
1\\
\hspace{1.2 in} \mathrm{for} \; \phi \le \phi_0, \\
\cos^2[(\pi / 2) (\phi- \phi_0) / \phi_1]\\
\\
\hspace{1.2 in} \mathrm{for} \; \phi_0 \le \phi \le \phi_0  + \phi_1, \\
0 \\
\hspace{1.2 in} \mathrm{for} \; \phi \ge  \phi_0  + \phi_1.\\
\end{array}
\right.
\end{equation} 
For $ss$ interactions, $\gamma_{orient} = \gamma_{orient}(\hat{\mathbf{r}}_{ij}, \mathbf
{\Omega}_i,\mathbf{\Omega}_j) = F(\theta_i;\theta_0,\theta_1) F(\theta_j;\theta_0,\theta_1) F(\psi_{ij}; 2\theta_0,2\theta_1)$, where $\mathbf{\Omega}$ describes particle orientation.  $\theta_i$ is the angle between the interacting patch on particle $i$ and $\hat{\mathbf{r}}_{ij}$, whilst $\theta_j$ is between the patch of particle $j$ and $-\hat{\mathbf{r}}_{ij}$. $\psi_{ij}$ is the angle between the projections of the membrane patches of $i$ and $j$ onto the plane perpendicular to $\hat{\mathbf{r}}_{ij}$. The factor $F(\psi_{ij})$ penalizes the twisting of interacting sub-units. We follow ref.~\cite{wilber} in choosing the range for this factor to be double that for the other ones. For $ms$ interactions, $\gamma_{orient} = \gamma_{orient}(\hat{\mathbf{r}}_{sm}, \mathbf{\Omega}_s) = F(\theta_s; \theta_0, \theta_1)$, where $\hat{\mathbf{r}}_{sm}$ is the unit vector from the sub-unit to the membrane particle, $\mathbf{\Omega}_s$ describes the orientation of the sub-unit and $\theta_s$ is the angle between the membrane patch and $\hat{\mathbf{r}}_{sm}$. We choose $\theta_1 = 0.2$ for all interactions.

To investigate the shape of assembled structures we employ the asphericity~\cite{aronovitz}:
\begin{equation}
\Delta = \frac{3}{2} \mathrm{tr} \, \mathbf{\hat{Q}}^2  / (\mathrm{tr} \, \mathbf{Q})^2 \hspace{0.5in} (0 \le \Delta \le 1),
\end{equation}
where $\mathbf{Q}$ is a $3\times3$ tensor whose elements are given by
\begin{equation}
\mathbf{Q}_{\alpha \beta} = \frac{1}{N}\sum_i r_{i,\alpha} r_{i,\beta}  - \frac{1}{N^2} \sum_i r_{i,\alpha} \sum_j r_{j,\beta},
\end{equation}
where the sums are over all the $N$ particles in the structure considered and $r_{i,\alpha}$ is one of the three cartesian components of the position of particle $i$. $\mathbf{\hat{Q}}$ is a traceless counterpart of $\mathbf{Q}$ defined as $\mathbf{\hat{Q}} = \mathbf{Q} - \bar{\lambda} \mathbf{I}$, where $\bar{\lambda}$ is the average of the three eigenvalues of $\mathbf{Q}$.

\clearpage

\subsection{Additional Results}

\begin{figure}[!h]
\begin{center}
\includegraphics[scale=0.8]{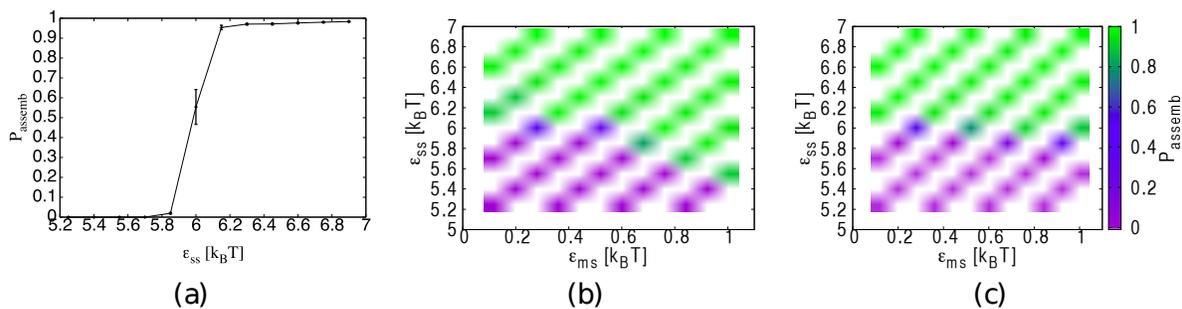}
\caption{Probability of finding a correctly assembled icosahedral core, $P_{assemb}$, in a simulation with 12 sub-units: (a) Without membrane as a function of $\epsilon_{ss}$. With the same free assembly volume as for the membrane simulations. Error bars show the standard error, calculated from 4 independent repeats of the simulation. Lines joining the data points added as a visual aid. With membrane, as a function of $\epsilon_{ss}$ and $\epsilon_{ms}$, for $\lambda_b$: (b) $2\sqrt{3} k_B T$ (c) $4\sqrt{3} k_B T$.}
\end{center}
\end{figure}

\begin{figure}[!h]
\begin{center}
\includegraphics[scale=0.6]{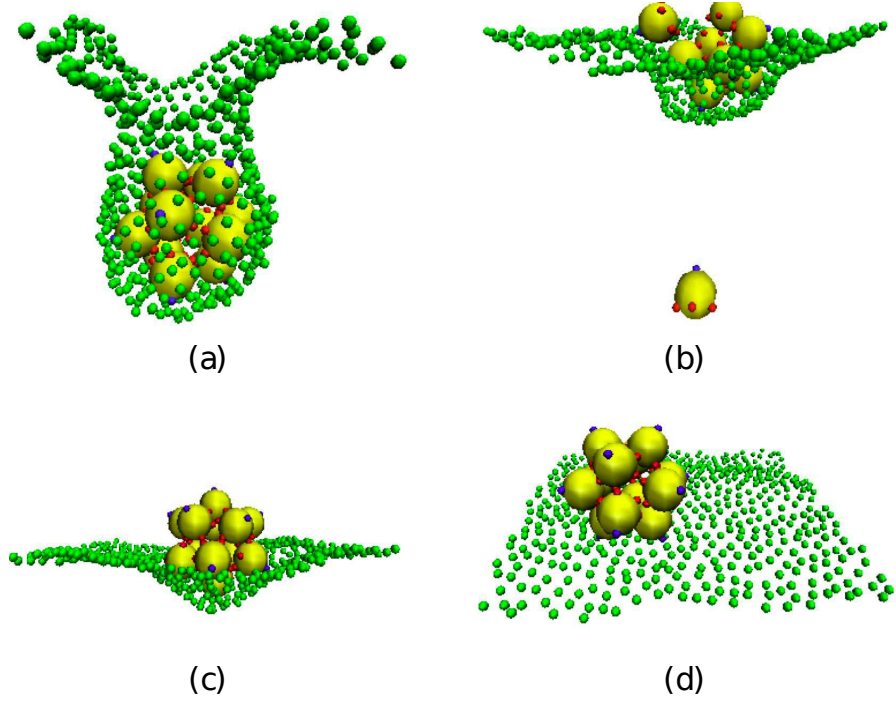}
\caption{Typical configurations for the core model for different $\lambda_b$: (a) Configuration showing an assembled, enveloped core for $\epsilon_{ss} = 5.55 k_BT$, $\epsilon_{ms} = k_BT$,  $\lambda_b =  2\sqrt{3} k_B T$. (b) Configuration from the system with the same parameter values as (a), showing a partially-assembled core that is also only partially enveloped. (c) Configuration showing  some deformation of the membrane for $\epsilon_{ss} = 6.90 k_BT$, $\epsilon_{ms} = k_BT$,  $\lambda_b =  4\sqrt{3} k_B T$. (d) Configuration showing no appreciable deformation of the membrane for $\epsilon_{ss} = 6.90 k_BT$, $\epsilon_{ms} = k_BT$,  $\lambda_b =  8\sqrt{3} k_B T$. Coloring as in the main text.}
\end{center}
\end{figure}

\begin{figure}[!h]
\begin{center}
\includegraphics[scale=0.6]{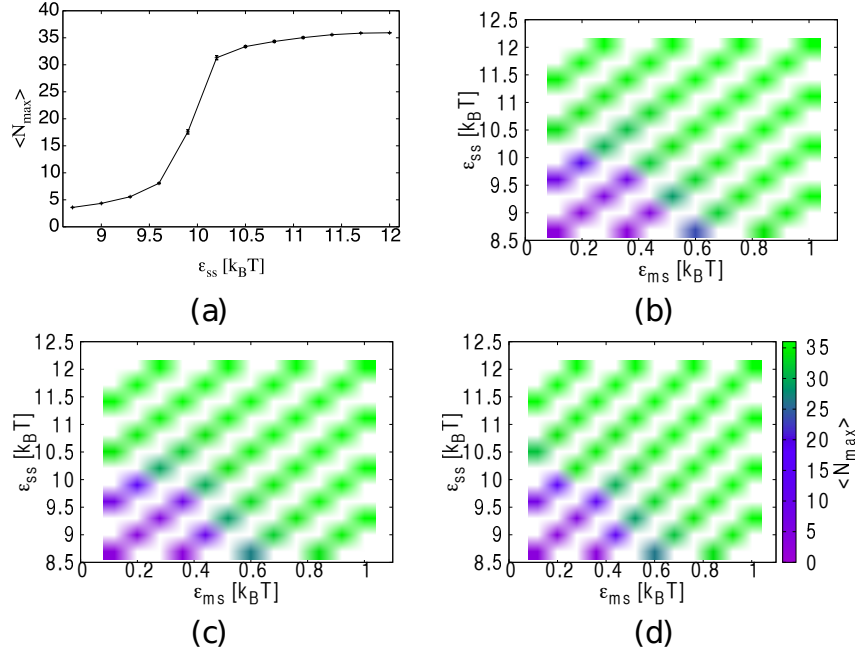}
\caption{Average number of sub-units in the largest bonded cluster, $\langle N_{max} \rangle$, for the clathrin-like model. (a) Without membrane, as a function of $\epsilon_{ss}$. With the same free assembly volume as for the membrane simulations. Error bars show the standard error, calculated from 4 independent repeats of the simulation. Lines joining the data points added as a visual aid. (b) As a function of $\epsilon_{ss}$ and $\epsilon_{ms}$, for $\lambda_b = \sqrt{3} k_B T$. (c) As a function of $\epsilon_{ss}$ and $\epsilon_{ms}$, for $\lambda_b = 2 \sqrt{3} k_B T$. (d) As a function of $\epsilon_{ss}$ and $\epsilon_{ms}$, for $\lambda_b = 4 \sqrt{3} k_B T$.}
\end{center}
\end{figure}

\begin{figure}[!h]
\begin{center}
\includegraphics[scale=0.45]{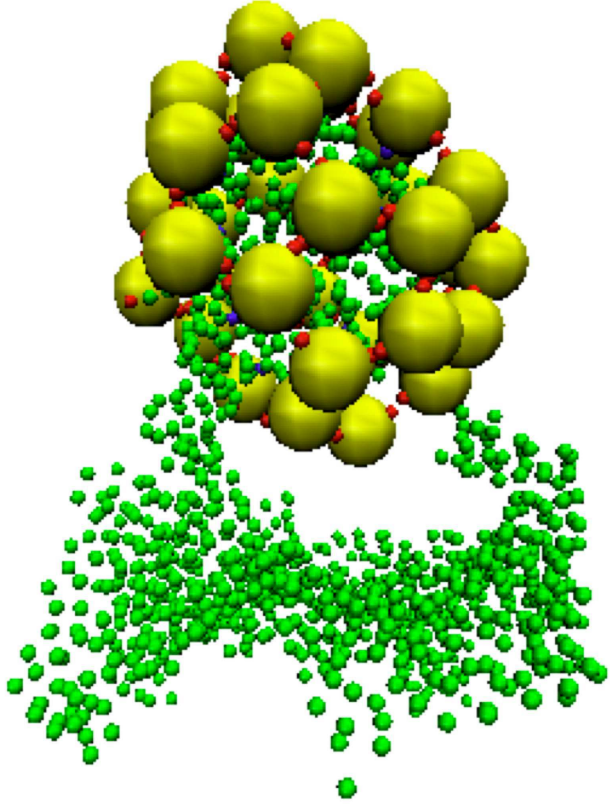}
\caption{Typical configuration for the clathrin-like model, for $\epsilon_{ss} = 12 k_BT$, $\epsilon_{ms} = k_BT$,  $\lambda_b =  \sqrt{3}/2 k_B T$, showing an almost-closed cage enclosing a region of membrane connected to the main part by a narrow neck (on the left of the image, the protrusion on the right is not directly connected to the enclosed region). Coloring as in the main text.}
\end{center}
\end{figure}

\clearpage

\end{document}